\documentstyle[aps,preprint]{revtex} 
\begin{document}

\def\half{{1 \over 2}}

\def\nfdf{${\cal N}=4, d=4$}
\def\ymt{Yang-Mills theory}
\def\sun{SU($N$)}
\def\laa{\langle\kern-.3em \langle}
\def\raa{\rangle\kern-.3em \rangle}
\def\la{\lambda}
\def\cn{{\cal N}}\def\cz{{\cal Z}}
\def\ymt{Yang-Mills theory}
\def\sftH{string field theory Hamiltonian}
\def\aaa#1{a^{\dagger}_{#1}}
\def\aap#1#2{(a^{\dagger}_{#1})^{#2}}
\def\ee{\hbox{e}}
\def\dd{\hbox{d}}\def\DD{\hbox{D}}
\def\tr{\hbox{tr}}
\def\Tr{\hbox{Tr}}
\def\part{\partial}
\def\dj#1{{\delta\over{\delta J^{#1}}}}
\def\cO{{\cal O}}
\def\cF{{\cal F}}
\def\cS{{\cal S}}
\def\cH{{\cal H}}
\def\cO{{\cal O}}
\def\cN{{\cal N}}\def\cL{{\cal L}}
\def\ssc{\scriptscriptstyle}
\def\refe#1{(\ref{#1})}

\title{Deformation quantization as the origin of D-brane 
non-Abelian degrees of freedom}
 
\author{Vipul Periwal}

\address{Department of Physics, 
Joseph Henry Laboratories,
Princeton University, 
Princeton, NJ 08544 
\tt vipul@princeton.edu}

\maketitle 
\begin{abstract}
I construct a map from the Grothendieck group of coherent sheaves to    
$K$-homology.  This  results in 
explicit realizations of $K$-homology cycles associated with D-brane
configurations.  Non-Abelian degrees of freedom arise in this 
framework from the deformation quantization 
of $N$-tuple cycles.  The large $N$ limit of 
the gauge theory on D-branes wrapped on a subvariety $V$ of some 
variety $X$ is geometrically 
interpreted as the deformation quantization of the formal completion 
of $X$ along $V.$
\end{abstract}
\bigskip
\tightenlines
\def\st#1#2{#1\star#2}
\def\poi#1#2{\{#1,#2\}}
\def\cf{Cattaneo-Felder }
\def\sw{Seiberg-Witten }

\section{Introduction}

Soon after Polchinski's identification of D-branes as the
nonperturbative objects in perturbative string theory that carry 
Ramond-Ramond charge\cite{p}, Witten\cite{wsym}\ argued that the low-energy 
effective action for the dynamics of nearly coincident D-branes should 
be appropriate dimensional reductions of supersymmetric Yang-Mills(sYM) 
theory in ten dimensions.  An exciting aspect of 
this identification\cite{wsym}\ was the fact that the transverse spacetime 
coordinates describing coincident D-branes metamorphosed into
Hermitian matrices, the adjoint scalars in the dimensional reduction 
of sYM theory.  Witten's argument was based on supersymmetry, 
specifically the BPS property of parallel D-branes. 
This appearance of non-Abelian structure  in spacetime co\"ordinates was 
the first indication in the context of a perturbatively consistent theory 
of quantum gravity that spacetime 
might have some intrinsic non-commutativity.

Following
seminal work by Sen\cite{sen}\  on the construction of non-BPS D-branes as 
solitons on the world-volumes of D-brane--anti-D-brane pairs, and 
other earlier work\cite{mm,cy}, 
Witten\cite{wk}\ suggested that D-brane charge should 
be classified by the compactly supported elements in the $K$-theory 
of spacetime.  Since $K$-theory deals with 
vector bundle equivalences, these constructions provide hints of 
non-Abelian degrees of freedom in D-brane physics, but no connection to
deformation quantization.   
On the other hand, Moore and Witten\cite{mw}\ argued 
that Ramond-Ramond fluxes should be classified by $K$-theory with an 
appropriate suspension of spacetime, an interpretation which received 
strong independent support\cite{dmw}.  

Finally, following pioneering work of Connes, Douglas  and 
Schwarz\cite{cds}, Seiberg and Witten\cite{sw}\  showed that 
non-commutativity appeared in open string theory via a deformation 
quantization interpretation of the effects of a large constant $B$ 
field on the worldvolume of a D-brane.  This non-commutativity seems 
to have nothing to do with the non-Abelian character of the
gauge theories relevant for D-brane physics mentioned above, though 
\cite{harvey}\ finds an intriguing connection in the presence of a 
large $B$ field.

This `historical' sketch
is an attempt to set the stage for the present 
paper.  I argued in \cite{meek} that D-brane charges should take values in
$K$-homology\cite{a,bdf,kas}\  instead of $K$-theory
essentially based on 
covariance with the original description of D-brane physics\cite{p}.
I suggested $KK$-theory as the appropriate framework for classifying 
configurations with both Ramond-Ramond charges and fields present.
 
The aim here is to show how non-Abelian  degrees of freedom 
arise from deformation quantization by linking D-brane charges to
$K$-homology\cite{meek}. 

\section{Outline}

In this section I outline the argument for the connection mentioned 
above.  Requisite details are given in later sections.

I first construct a map from the Grothendieck $K$-group of coherent 
sheaves to $K$-homology.  (Sharpe\cite{sharpe}\ was the first to note the
relevance of the $K$-theory of coherent sheaves in the context of D-branes.)  
The first fact 
needed is that coherent sheaves are pushed-forward under morphisms of 
varieties.   The coherent
sheaf $K$-groups, which we denote $G^{*}(A),$ where $A$ is the ring of 
interest, transforms covariantly under morphisms of varieties and as 
such we expect that there should be a map from $G^{*}(A)$ to the
$K$-homology of the variety, just as there is a dual map from the 
$K$-theory of locally free sheaves (equivalent to algebraic or 
analytic vector bundles) to the topological $K$-theory.  
I have not been able to find a direct treatment of this map from coherent 
sheaves to $K$-homology cycles 
in the mathematics literature, so I will give  a direct construction 
later.  The map from projective resolutions of structure sheaves of 
subvarieties to $K$-theory is, of course, well-known.

The important point about coherent sheaves is that they are supported 
on algebraic subvarieties.  Thus, for example, on the real line
considering the ring of polynomials $R$ and the ideal of functions 
that vanish at $x=0$ to order $N,$ $R/(x^{N})$ is a coherent sheaf on 
the real line which vanishes everywhere except for the stalk over
$x=0.$  This brings us to our second fact: Merkulov\cite{merk}\ showed 
that the deformation quantization of the function algebra associated 
to this `$N$-tuple point' is the algebra of  $N\times N$ matrices.  
The deformation parameter of physical interest is real, not imaginary
as is the  mathematicians' wont, which results in the appropriate 
algebra being the complex $N\times N$ matrices, with the physically 
appropriate real section being that of Hermitian matrices.  

Since the appearance of Merkulov's paper, it has been a puzzle as to 
how this calculation fits in with D-brane physics, since there was 
no context in which D-branes were naturally described in terms of 
algebraic equations defining subvarieties.  Now,
finally, we bring these facts together by noting that if, as argued in 
\cite{meek}\ on grounds of covariance, D-brane charge is classified 
by $K$-homology, there is a direct link between the deformation 
quantization of coherent sheaves associated to $K$-homology cycles and 
the appearance of non-Abelian degrees of freedom on D-brane 
worldvolumes.  I should note that this is not the  
application suggested in \cite{merk}.

According to the construction given here, there is a 
canonical way of computing the operator algebra associated with a given 
D-brane configuration on any algebraic variety.  The matrix model 
description of intersecting branes on a variety is an especially 
interesting special case\cite{gukov}.
In the case that D-brane worldvolumes are non-compact (as is usual 
for D$p$-branes ($p\ge 0$) in Minkowski space, for example), a more 
appealing formalism for describing D-branes is in terms of
a family (parametrized by the worldvolume) of transverse $K$-homology 
cycles.  The geometry of the 
large $N$ limit is that of a deformation quantization of the 
formal completion 
of the ambient variety along the support of the coherent sheaf.  
These topics are discussed in the last section.

Now for the details with no pretence at mathematical sophistication.

\section{Coherent sheaves to $K$-homology}

\cite{sharpe}\ reviewed 
the Grothendieck 
constructions in the context of $K$-theory and D-branes\cite{wk}, 
but the emphasis was 
on smooth projective varieties where Poincar\'e duality gives an isomorphism
between the groups $K_{*}(A)$ and $G^{*}(A).$  In the physically 
interesting situation of non-compact manifolds there is no canonical
isomorphism between these groups.   
The non-compact case leads to   another motivation for identifying 
D-brane charge with $K$-homology: Recall 
that the homology of singular chains and de Rham cohomology with compact 
supports are isomorphic, and that the duals of  compactly supported 
de Rham cohomology groups are isomorphic to de Rham cohomology 
groups\cite{bott}\ without  assumptions of finiteness or 
compactness.  According to Witten\cite{wk}, D-brane charge is also 
most naturally measured in terms of compactly supported (in 
transverse directions) $K$-theory 
classes, so one can again argue for $K$-homology as the appropriate 
group.  

Before giving the construction of $K$-homology cycles from coherent 
sheaf data, I recall the definition of these cycles\cite{kas,higson}.  
(A more compact form of the following 
definition would be in terms of ${\bf Z}_{2}$ graded
objects.)  A cycle in $K^{0}(A),$ the $K$-homology group of 
the algebra $A,$ is specified by two Hilbert spaces $\cH_{0,1},$ a 
linear map $\Theta:\cH_{1} \rightarrow \cH_{0},$ and representations 
$\phi_{i}$ of
$A$ on $\cH_{i}, i=1,2.$  These are linked by requiring that 
for any $a$ in $A,$  the operators
\begin{equation}
    \phi_{0}(a)\Theta - \Theta\phi_{1}(a), \quad
\phi_{0}(a)(\Theta\Theta^{*}-1),\quad {\rm  and} \quad\phi_{1}(a)
(\Theta^{*}\Theta-1)
\label{require}
\end{equation}
are all compact 
operators.  For orientation, on a non-compact manifold $X$
a possible choice for $A$ is the algebra of continuous bounded 
functions that vanish at infinity, $C_{0}(X),$ and the Hilbert spaces 
could be the spaces of $L^{2}$ sections of two vector bundles 
$E_{i},i=0,1$
on $X,$ with the obvious action of continuous functions on sections of
vector bundles by pointwise multiplication.  It is important in this 
non-compact context to notice that we are not assuming that $\Theta$ is
Fredholm.

A coherent sheaf is a sheaf that has 
a projective resolution by locally free sheaves\cite{gunning}.  
This means there is an exact sequence
\begin{equation}
    \ldots\rightarrow \cF_{n}\rightarrow \ldots \cF_{0} \rightarrow 
    \cS \rightarrow 0
    \label{coherent}
\end{equation}
where the sheaves $\cF_{i}$ are sheaves of  sections of vector bundles.  
Coherent sheaves over varieties therefore have the 
possibility of being compactly supported, since the dimension of the 
stalks of $\cS$ is related to the rank of the linear maps between 
$\cF_{i}.$  We therefore compute the values of 
determinants of submatrices of 
these maps, and the zeros of these determinants determine the support 
of $\cS.$  (Coherent sheaves can be defined much more
generally, not just on analytic or algebraic varieties.  Their 
description in terms of finitely generated modules defined by a 
finite set of relations is appropriate in general\cite{grauert}.)

An alternative local characterization of a coherent sheaf that we will 
use  is the following:  Locally a coherent sheaf has a projective 
resolution by free modules
\begin{equation}
    \cO^{n} \mathrel{\mathop\rightarrow^{{\scriptstyle\sigma}}} 
    \cO^{m} \rightarrow \cS \rightarrow 0
    \label{coherentlocal}
\end{equation}
where $\cO$ is the sheaf of (germs of) holomorphic functions.
These local resolutions 
do not necessarily fit together globally, hence the need for the 
general characterization given in \refe{coherent}.  Now, by extending 
with the identity map, we can take \refe{coherentlocal}\ to be of the 
form
\begin{equation}
    \cO^{\infty} \mathrel{\mathop\rightarrow^{{\scriptstyle\hat\sigma}}} 
    \cO^{\infty} \rightarrow \cS \rightarrow 0
    \label{coherentinf}
\end{equation}
where we assume that $\hat\sigma -1$ is of finite rank.  

We now go out of the algebraic or   analytic
category and consider a partition of 
unity subordinate to these local projective resolutions 
\refe{coherentinf}.   Since we want to construct a map to 
$K$-homology, we expect to lose the algebraic structure in any case,
so this is acceptable.  By 
polynomial approximation in each patch, we extend $\hat\sigma$ to 
smooth sections.  Finally we recall that Hilbert bundles are 
trivial since the infinite unitary group is contractible so we 
encounter no obstructions in patching these local projective 
resolutions together to obtain a global resolution of $\cS$ by
sheaves of sections of Hilbert bundles $E_{i}$ 
\begin{equation}
    \Gamma(E_{1}) \rightarrow \Gamma(E_{0}) \rightarrow \cS \rightarrow 
    0 ,
    \label{coherenthilb}
\end{equation}
picking arbitrary Hermitian metrics on the fibres.
We restrict ourselves now to a coherent  sheaf $\cS$ 
such that the support of $\cS$  
is strictly contained in the variety $X.$  Then $\hat\sigma$ extends 
to a map $\Theta$ from the Hilbert space of $L^{2}$ sections of $E_{1}$ to 
the Hilbert space of $L^{2}$ sections of $E_{0} $ which is almost 
everywhere the identity.   Since the whole 
construction is local, the action of  $ \Theta$ commutes with
that of pointwise multiplication by functions in $C_{0}(X).$  Finally 
while we do not expect $\Theta\Theta^{*}-1$ or its adjoint to be compact for 
$X$ non-compact, we do expect \refe{require}\ to be satisfied with the 
restriction that the support of $\cS$ is strictly contained in $X.$

It seems straightforward to argue that the map from coherent sheaf 
data to $K$-homology data we have just constructed is well-defined in
going to $K$-homology since we allow ourselves the use of smooth homotopies
even though  a short exact sequence of 
coherent sheaves 
\begin{equation}
    0 \rightarrow \cS' \rightarrow \cS \rightarrow \cS'' \rightarrow 0 
\end{equation}
does not necessarily split algebraically or analytically.
Thus we have finished constructing a map from coherent sheaves to
$K$-homology, and can therefore think of $K$-homology cycles as being 
associated with  subvarieties  in $X$ that arise  as the support of 
coherent sheaves.  This is more data than just specifying a subvariety
as a homology cycle.

\section{Deformation quantization of coherent sheaves}

I turn now to a brief recapitulation of Merkulov's 
calculation\cite{merk}.  Consider the $N$-tuple point, the solution of the 
equation $x^{N}=0$ on the affine real line.  The cotangent bundle has 
a canonical symplectic structure so the complexified 
algebra of functions on this
phase space, extended to formal power series in the 
deformation parameter, can be deformed via the Moyal product\cite{moyal}:
\begin{equation}
    f*g(x,p) \equiv \exp\left({i\hbar\over 2}\left[{\part^{2}\over{\part 
    p\part \tilde x}} -{\part^{2}\over{\part 
    \tilde p\part  x}} \right]\right) f(x,p)g(\tilde x,\tilde 
    p)|_{(x,p)=(\tilde x,\tilde p)}.
\end{equation}
The obvious projection $\pi$ from the phase space to the real line 
leads to a pullback of the ideal generated by $x^{N}$ in the 
algebra of functions on the line.  The ideal $I_{N}$ in the deformed algebra
consists of elements of the form $g(x,p)* \pi^{*}(x^{N}) .$  The 
normalizer $\cN_{N}$ of this ideal is  
\begin{equation} 
    \cN_{N}\equiv \{f(x,p) : \pi^{*}(x^{N})*f\  {\rm is}\ {\rm in} \ 
    I_{N} \} .
    \label{normalize}
    \end{equation}
The point of introducing the 
normalizer is that $I_{N}$ is a two-sided ideal inside $\cN_{N}.$  We 
can therefore construct the quotient algebra.  Finally, 
by explicit computation, Merkulov showed that $\cN_{N}/I_{N}$ is 
isomorphic to the algebra of $N\times N$ complex matrices, with the 
induced product in the quotient algebra mapping into matrix 
multiplication.  

A remarkable fact about this construction is that the 
induced product is meromorphic in $\hbar.$  The reason for this is 
interesting, since na\"\i vely one might have supposed that the 
functions on phase space would be generated by monomials of the form
$x^{i}p^{j}, 0\le i,j \le N-1.$  This is not however the case and 
there are terms with higher powers of $p$ present.  For example, the 
general element in the quotient algebra $\cN_{2}/I_{2}$ is\cite{merk}
\begin{equation}
    h \equiv a+ b( p + {i \over \hbar}  p^{2}  * x)
    + (c + dp) * x,
    \label{neqtwo}
\end{equation}
with $a,b,c,d$ complex numbers.
By dimensional analysis, the higher power  of $p$ in \refe{neqtwo},
$bp^{2}*x$ must be   accompanied 
by an appropriate negative power of $\hbar.$  
In other words, the $N$-tuple point is a well-behaved quantum system 
with no classical phase space Poisson algebra.  The reader surprised 
by the appearance of terms singular as $\hbar\downarrow 0$ may wish 
to look at other examples\cite{me}. 

Merkulov's construction is much more general than the example I have
reviewed.  As I mentioned earlier, the difficulty in utilizing this 
construction in matrix models of D-brane dynamics arises because {\it 
a priori} there is no reason to associate D-branes with solutions 
of algebraic equations.  However, in light of our association of
D-brane charge with $K$-homology, and the construction above of
$K$-homology cycles from coherent sheaves, we see  a natural 
application of Merkulov's idea.  Namely, the deformation quantization 
of the coherent sheaf associated with a $K$-homology cycle is the 
operator algebra of interest. 

$K$-theory is invariant under deformation quantization\cite{rosen}, 
a fact used in
\cite{sergei}\ to understand D-brane phase transitions in $K$-theory.
It is reasonable to expect then that the homology theory dual to $K$-theory,
{\it i.e.} $K$-homology, will also be invariant.  

To formalize the deformation quantization version of the $K$-homology 
cycle, we now take $\cH_{i}\equiv L^{2}(T^{*}{\bf R})\cap \cN_{N}, 
i=0,1$
and the operator $\Theta$ to 
be an appropriate bounded version of multiplication by  $x^{N},$ for 
example:
\begin{equation}
    \Theta:f(x,p) \mapsto 
\pi^{*}(x^{N}/(1+x^{2N})^{{1/2}})*f(x,p).
\end{equation}
(The definition of $\cN_{N}$ in \refe{normalize}\ 
needs appropriate modification of course.)
The algebra of interest in this case is $A\equiv C_{0}({\bf 
R}).$  We need to check that the representation of this algebra on
$\cN_{N}$ is actually a representation, in other words for all
$a$ in $A$ and $f$ in $\cN_{N},$  $\pi^{*}(a) * f $ is in $\cN_{N}.$
This is trivially verified using the associativity of the 
$*$ product and the definition of $\cN_{N}.$
Finally this representation of $A$ on $\cH_{i}$  
commutes  with the operator $\Theta.$

\section{Superconnections, families and formal completions}

There is, of course, more to D-branes
than just the appearance of matrix degrees of freedom.  It is easy to
see that an $N$-tuple point on a plane, for example, will lead to an
algebra of two independent matrices and so on. 
Thus co\"\i ncident  D(-1)branes are easily described in this framework.
Further, for a D-brane wrapped on a cycle in a compact variety, we 
get an operator algebra description that is not described in terms of 
matrices parametrized by a worldvolume, but instead directly in terms 
of an operator algebra {\it including} the worldvolume\cite{gukov}.

For a non-compact worldvolume, it is more appropriate to think of a 
family, parametrized by the worldvolume, of operator algebras
describing the transverse geometry.  The operator $\Theta$ is now 
parametrized by the worldvolume and is combined in a superconnection 
with a connection $\nabla : \cH_{i} \rightarrow \cH_{i}.$ 
It is a great deal simpler to work throughout with a
${\bf Z}_{2}$-graded formalism as set out in \cite{cayley}.  This
would lead us quite far from the main point of this paper so  I will 
address this elsewhere.

I want to briefly address perhaps the most interesting aspect of
this formalism, which  is to relate the coupling of Ramond-Ramond fields to
D-brane worldvolumes to the pairing between $K$-theory and $K$-homology.
Just as $K$-theory classes are represented as cohomology classes via 
the Chern homomorphism, algebraic $K$-theory classes are represented 
as cyclic homology cycles and $K$-homology cycles are represented via
the dual Chern homomorphism as cyclic cohomology cocycles\cite{connes}.  
To get an 
intuitive idea for what is involved in this, recall that 
\begin{equation}
    \cL_{v} = \dd \iota_{v} + \iota_{v} \dd .
\end{equation}
The Chern-Simons coupling as usually described in the 
literature\cite{rob}\ takes 
the form 
\begin{equation}
\int_{V} \tr \exp(F + \iota_{\phi}^{2}) \phi^{*}C
\label{chern}
\end{equation}
where (abusing notation) 
$\phi^{*}C$ is the pullback of the spacetime Ramond-Ramond potential to 
the worldvolume of the D-branes with $\phi$ describing the fluctuation 
of the D-brane worldvolume in 
the transverse directions (in the absence of a $B$ field).  
Computing $(D + \iota_{\phi})^{2},$ where $D=\dd + A,$ we find
\begin{equation}
    (D + \iota_{\phi})^{2} = F + \iota_{\phi}^{2} + \hat\cL_{\phi},
\end{equation}
where $\hat\cL$ is a gauge covariant Lie derivative.  
The integral in \refe{chern}\ is over the worldvolume $V$ so we can 
try to  think of it as
\begin{equation}
    \int_{V} \tr\exp(F + \iota_{\phi}^{2}) \exp(\hat\cL_{\phi}) C
    \label{chernpullback}
\end{equation}
and hence in the form
\begin{equation}
    \int_{V} \tr\exp((D+ \iota_{\phi})^{2}) C ,
    \label{coupling}
\end{equation}
where we have ignored the fact that the operators involved 
do not commute.
Nevertheless, the correct form of \refe{coupling}\ appears to be
the JLO-cocycle as represented  in \cite{quillen}.
 
Finally, a geometric picture of the large $N$ limit emerges from
the representation of D-branes given in this paper.  Recall a standard 
construction in algebraic  geometry:  If a subvariety is defined by
the vanishing of some ideal of functions $I$ then 
the $N^{\rm th}$ infinitesimal 
neighbourhood of a subvariety is defined by the vanishing of $I^{N}.$
The formal completion\cite{hartshorne}\
of these infinitesimal neighbourhoods is 
analogous to the construction of $p$-adic numbers from the integers.
Thus, the geometric meaning: The large $N$ limit of D-brane 
physics is  the deformation quantization of the formal
completion.  
\bigskip

Acknowledgements: 
I am indebted to S. Gukov and R.C. Gunning for  enlightening discussions.
In particular,  Gukov pointed out the relevance of Sharpe's work. 
I am grateful to G. Moore and J. Rosenberg for helpful remarks.
This work was supported in part by NSF grant PHY98-02484.

\def\np#1#2#3{Nucl. Phys. B#1,  #3 (#2)}
\def\prd#1#2#3{Phys. Rev. D#1, #3 (#2)}
\def\prl#1#2#3{Phys. Rev. Lett. #1, #3 (#2)}
\def\pl#1#2#3{Phys. Lett. B#1, #3 (#2)}
\def\jhep#1#2#3{J. High Energy Phys. #1, #3 (#2)}

\end{document}